\documentclass[journal]{IEEEtran}
\usepackage{multirow}
\usepackage{footnote}

\usepackage{amsmath,amssymb,amsfonts}
\usepackage{algorithmic}
\usepackage{graphicx}
\usepackage{textcomp}
\usepackage[export]{adjustbox}
\usepackage{xcolor}
\usepackage{gensymb}

\usepackage{listings}

\definecolor{codegreen}{rgb}{0,0.6,0}
\definecolor{codegray}{rgb}{0.5,0.5,0.5}
\definecolor{codepurple}{rgb}{0.58,0,0.82}
\definecolor{backcolour}{rgb}{0.95,0.95,0.92}

\lstdefinestyle{mystyle}{
	commentstyle=\color{codegreen},
	keywordstyle=\color{magenta},
	numberstyle=\tiny\color{codegray},
	stringstyle=\color{codepurple},
	basicstyle=\ttfamily\footnotesize,
	breakatwhitespace=false,         
	breaklines=true,                 
	captionpos=b,                    
	keepspaces=true,                 
	showspaces=false,                
	showstringspaces=false,
	showtabs=false,                  
	tabsize=2
}

\title{Compact Modeling and Rapid Simulation of Silicon Photonic Micro-Disk and Ring Modulators}

\author{Vishal~Saxena and~Md Jubayer Shawon % <-this % stops a space
\thanks{V. Saxena and Md Jubayer Shawon are with the ECE Department, University of Delaware, Newark,
DE. E-mail: vsaxena@udel.edu}% <-this % stops a space
}

\begin{document}

\maketitle



\begin{abstract}
Microdisk or microring modulators (MDMs or MRMs) realize compact electro-optic modulation in active silicon photonics (SiP) foundry platforms. A key advantage of these resonant modulators is that they readily implement dense wavelength division multiplexing (DWDM) in optical interconnects by tuning and locking the MDMs/MRMs to the DWDM wavelength grid. Compact modeling of static and transient dynamics of these modulators is important for co-simulation with CMOS drivers and wavelength stabilization circuits. This work presents the first compact model for microdisk modulators with a novel approach that uses experimental measurements and allows rapid and accurate simulation.  This approach employs coupled real-valued differential equations with analytic signals in Verilog-A, leading to a 7X speed up in transient simulation time over the current art while enhancing accuracy. Since the model is generalized, it can also model microring resonators in addition to MDMs. The model also includes thermo-optic tuning for MDM/MRM with an embedded microheater, which is essential for simulations involving resonant wavelength stabilization.
\end{abstract}


%%%%%%%%%%%%%%%%%%%%%%%%%%  body  %%%%%%%%%%%%%%%%%%%%%%%%%%
\section{Introduction}

Resonant modulators realize compact and lumped electro-optic modulation in silicon photonic (\textbf{SiP}) based photonic integrated circuits (\textbf{PIC}s). These microdisk modulators (\textbf{MDMs}) and microring modulators (\textbf{MRMs}) are promising for several applications, including on-chip and data center interconnects \cite{li201525,wade2021monolithic,saxena2024ringTcas2}, optical switches \cite{rizzo2022petabit}, and recently optical neural network accelerators \cite{huang2022prospects} and programmable photonics \cite{shawon2023silicon}. A key advantage of these resonant modulators is that they incur a compact layout footprint and readily implement dense wavelength division multiplexing (\textbf{DWDM}) by locking them to the DWDM wavelength grid \cite{ding2014compact,sharma2021silicon,rizzo2022petabit}.  Co-design of these resonant modulators with CMOS driver circuits is essential for pre-silicon validation of the optical transmitters \cite{mishra2022hybrid}. This necessitates a compact model that enables rapid simulation even at very high data rates exceeding 25Gbps. Verilog-A-based models are being prolifically used to co-simulate optical components with CMOS circuits in Cadence or similar electronic design automation (\textbf{EDA}) environment \cite{kehan_MW2014,zhuTCAS2015,bcicts2018,shawon2020rapid,shawon2022JLT,shawon2022JLT,shawon2023silicon,shawon2023automatic}. 

While passive optical circuits and Mach-Zehnder modulators (MZMs) have been extensively modeled using Verilog-A \cite{kehan_MW2014,zhuTCAS2015,zhu2016modeling}, MRM models have lagged behind. Early attempts employed composite MRM modeling approach \cite{sorace2015electro}. These models incur long simulation times and deviate from experimental measurements. Lumped models were introduced for MRMs \cite{rhim2015verilog,wang2016compact}, which are the current art. However, these models employed approximated differential equation solutions, did not model the voltage-dependent electrical effects, and used a reference clock, which incurs slower simulation time. 
In this work, we alleviate the limitations of current art, resulting in a comprehensive and accurate compact model for depletion-mode resonant modulators. This is the first experimentally extracted compact model for microdisk modulators in the literature. Since the model is generalized, it can also simulate MRMs in addition to MDMs. Another novelty of this compact model is that it employs a coupled real-valued analytic differential equations in Verilog-A that leads to an impressive 7X speed up in transient simulation time over the current art while capturing the dynamic nonlinear effects observed experimentally with higher fidelity. Finally, the third novelty is that the model also includes thermo-optic tuning for MDM/MRM with an embedded microheater, which is essential for simulations involving resonant wavelength stabilization of these modulators. 

%%%%%%%%%%%%%%%%%%%%%%%%%%%%%%%%%%%%%%%%%%%%%%%%%%%%%%%%%%%%%%%%%%%%%%%%%%%%%%%%%%%%%%%%%%%%%%%%%%%%%%%%
 \section{SiP Microdisk Fabrication}
% In this section we describe the device and model fitting process used to experimentally verify our model.  

\subsection{Foundry SiP Process}
\begin{figure}[!ht]
\centering
\includegraphics[width=\linewidth]{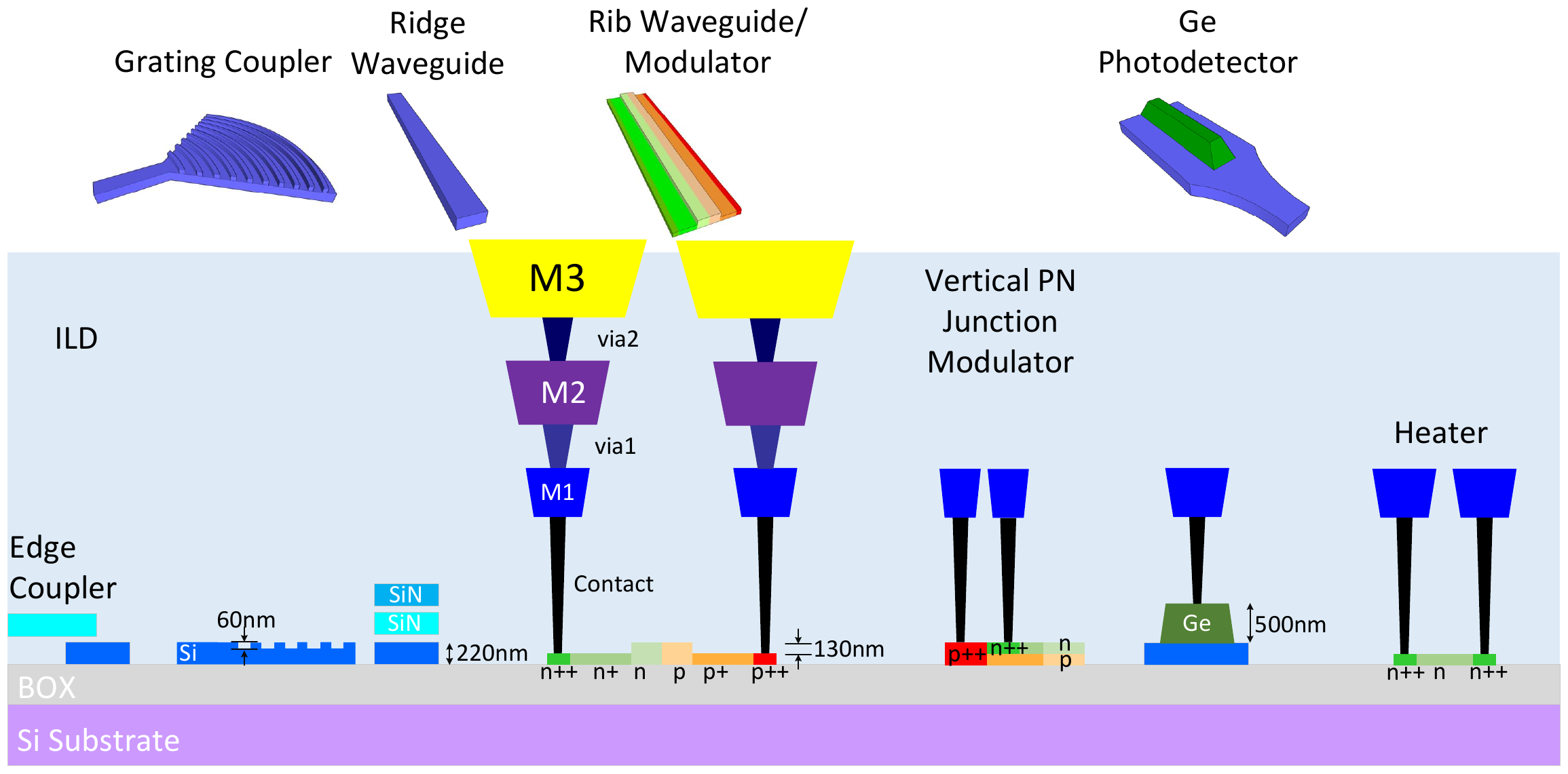}
\caption{Cross-section of the SiP process featuring SOI and nitride waveguides, doping options, Ge detectors, and metal layers for routing \cite{fahrenkopf2019aim}. 
%(b) Chip micrograph of the fabricated microdisk modulator. 
}
\label{fig:AIMprocess}
\end{figure}

The resonant modulator test structures used in this work were fabricated in the AIM Photonics multi-project wafer (MPW) run using the AP Suny PDK components \cite{fahrenkopf2019aim}. A cross-section of the AIM foundry process is shown in \textbf{Fig.} \ref{fig:AIMprocess}(a). The process employs 300mm silicon-on-insulator (\textbf{SOI}) wafers with thick buried oxide (BOX), 220nm SOI layer for silicon waveguides, and etched silicon thickness of $\sim$110 nm.  The process uses 193 nm immersion lithography with critical dimensions down to 100 nm and features several n- and p- doping levels and depths to realize lateral and vertical \textit{pn}-junctions. While this process employs doped silicon microheaters, few other foundries provide metal heaters for lower voltage operation.

\subsection{Microdisk Modulator}
\begin{figure}[!h]
\centering
\includegraphics[width=\columnwidth]{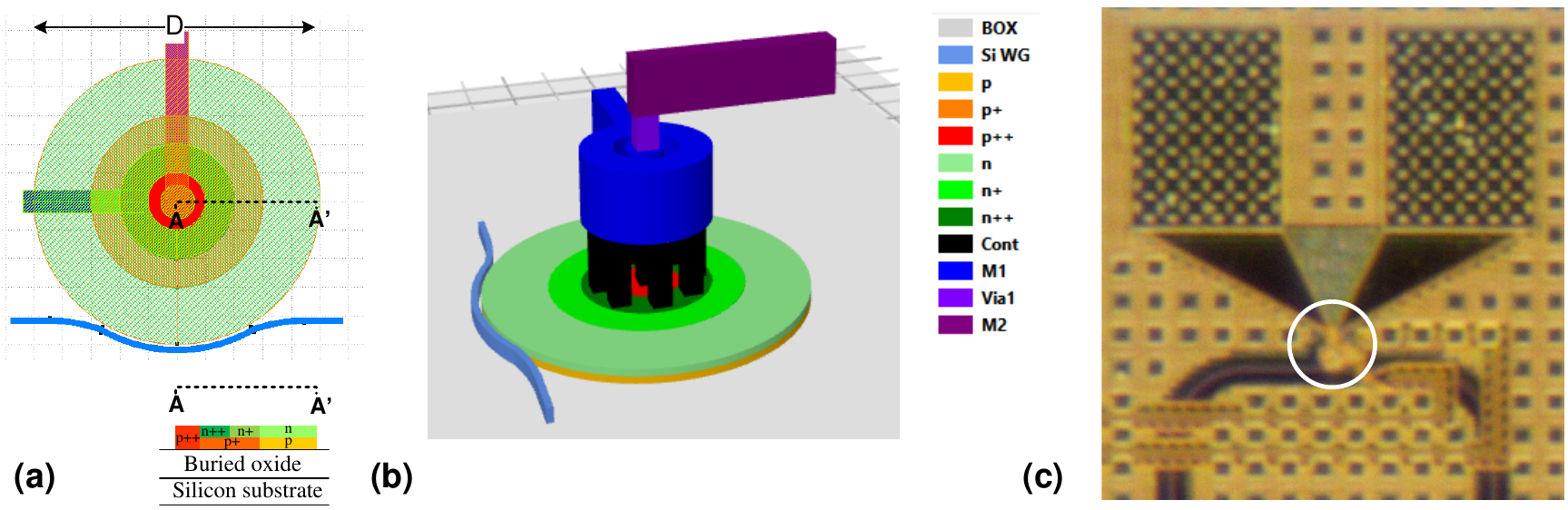}
\caption{(a)  Layout of a microdisk modulator (MDM) in a silicon photonic process, (b) 3D view, (c) fabricated chip micrograph.}
\label{fig:Ring_Modulator_v1} 
\end{figure}

An MRM or MDM is constructed using a silicon-based \textit{pn}-junction microring or microdisk structure, respectively. MRMs are fabricated as a single-mode rib waveguide with lateral or interleaved \textit{pn}-junction in the form of a ring coupled with a bus. MRRs have been designed with a radius in the 5-30$\mu m$ range \cite{baehr201225,pantouvaki201556gb,saxena2024ringTcas2}. On the other hand, microdisk modulators exploit the overlap of the whispering gallery mode in a very small ($<$5$\mu$m) disk with a vertical \textit{pn}-junction \cite{timurdogan2013vertical,timurdogan2014ultralow,rizzo2022petabit}. \textbf{\textit{Fig.}} \ref{fig:Ring_Modulator_v1} shows the simplified structure of a microdisk modulator. Exploiting the small disk radius, MDMs exhibit a higher free-spectral range (\textbf{FSR}) than MRMs, making them amenable to DWDM. The higher mode-overlap with junction modulation in a vertical MDM and a higher quality factor (Q) due to lower loss enables sub-1V drive voltages.

The compact C-band MDM has a diameter ($D=2R \sim 5\mu m$) and is realized using a depletion-mode (i.e., reversed bias) pn-junction with an integrated silicon heater \cite{fahrenkopf2019aim}. The MDMs in the literature typically employ full silicon etch to form the disk and a vertical pn-junction to maximize the optical mode overlap with the pn-junction \cite{rizzo2022petabit}. Due to their higher optical confinement and mode overlap with the change in carrier concentrations, MDMs realize higher modulation efficiency compared to the microring modulators \cite{timurdogan2013vertical,timurdogan2014ultralow}.  
The MDM in the AIM APSUNY PDK used in this work exhibits $\sim 8$ GHz/V detuning efficiency and is capable of 25 Gbps NRZ data rates with over 4dB extinction ratio and sub $1V_{pp}$ drive voltage \cite{fahrenkopf2019aim}. The embedded heater in the device operates with up to 8V \cite{fahrenkopf2019aim}. In our PIC seen in \textbf{Fig.} \ref{fig:AIMprocess}(b), the MDM is accessed through on-chip grating couplers, and additional monitor taps were included to assist a thermo-optic wavelength stabilization feedback loop \cite{li202012,kumar2023power}.

%%%%%%%%%%%%%%%%%%%%%%%%%%%%%%%%%%%%%%%%%%%%%%%%%%%%%%%%%%%%%%%%%%%%%%%%%%%%%%%%%%%%%%%%%%%%%%%%%%%%%%%%
\section{Resonant Modulator Modeling}

Modeling the optical response is challenging due to the modulation of the ring resonance and its quality factor with applied voltage and the photon lifetime in the ring. The optical behavior can be modeled using its components, \textit{i.e.} waveguides, \textit{pn}-junction phase modulator, and coupler using composite modeling approach \cite{sorace2015electro, RingModLumerical}. This composite model poses difficulty in transient simulations as the MRM resonant frequency is not directly known and needs to be determined using slow and time-consuming stepped-frequency transient analysis (SFTA) or frequency-chirp method (FCM) simulations \cite{shawon2019rapid, shawon2020rapid}. Also, the simulated response doesn't exactly match the experimental results due to process variations and unaccounted optical loss.

% \begin{figure}[!h]
% \centering
% \includegraphics[width=0.5\columnwidth]{figures/resonator1_heater}
% %\includegraphics[width=\columnwidth]{figures/MRR_heater_sch.png}
% \caption{(top) A ring resonator model with a thermal phase shifter. (bottom) Verilog-A modeling of the MRR in Cadence Virtuoso using our custom SiP PDK components.}
% \label{fig:resonator1_heater} 
% \end{figure} 

\begin{figure}[!ht]
\centering
\scriptsize
\includegraphics[width=\linewidth]{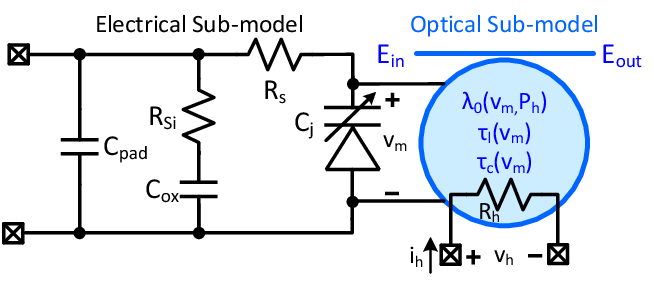}
\caption{The micro-ring/disk compact model showing the electrical and optical sub-models.}.
\label{fig:MDM_model}
\end{figure}

Lumped modeling is better suited for numerically fitting experimentally measured response of the resonant modulators to their compact model \cite{rhim2015verilog,zhang2010silicon,little1997microring,li201525}. The MRM/MDM compact model comprises electrical and optical sub-models as shown in \textbf{Fig.} \ref{fig:MDM_model}. The electrical model captures the pad capacitance, depletion-mode pn-junction capacitance, $C_j$, series resistance, $R_s$, and pad and substrate parasitics. These electrical parameters are obtained by fitting the measured $S_{11}$ parameter response to the model seen in \textbf{Fig.} \ref{fig:static_response} \cite{rhim2015verilog} and discussed later in \textbf{Section} \ref{sec:model_fitting}. 

\subsection{Optical Lumped Modeling} \label{sec:optical_lumped_modeling}
%It utilizes coupled-mode theory which, instead of the electric field, uses the energy amplitude in the resonant device \cite{little1997microring,zhang2010silicon}.
%The ring/disk supports a traveling wave of amplitude $A(t)$, which is normalized to make $|A(t)|^2$ equivalent to power flowing through the ring cross-section. 
In this approach, The modulator is modeled as a lumped resonator of energy amplitude $a(t)$, normalized to let $|a(t)|^2$ represent the net energy in the ring \cite{little1997microring,zhang2010silicon}. This approach generalizes the dynamics of MRM as well as the MDM. The resonant frequency is given as $\omega_0=\frac{2\pi c}{\lambda_0}=\frac{2\pi mc}{nL}$, where $n$ is the effective index, and $L$ is the effective circular length of the resonator core. Here, $c$ is the speed of light in free space, and $m$ is the resonance order. 
The resonator amplitude decays with a net time-constant, $\tau$, is contributed by the total losses in the ring/disk due to optical losses ($\tau_l$), and the power leaving into the coupled bus ($\tau_c$) \cite{little1997microring}. The net time-constant, $\tau$, and is given by 

% which in our analytic framework, can be substituted by $\omega_0 \rightarrow \omega_0-\omega_R$. 

\begin{equation} \label{eq:net_tau}
	\frac{1}{\tau} = \frac{1}{\tau_c} + \frac{1}{\tau_l}
\end{equation}

The energy amplitude in a single-bus resonant modulator, $a(t)$, is described by the differential equation \cite{little1997microring}

%\begin{align} \label{eq:ring_partial}
%    & \frac{\partial a(t)}{\partial t} = \Big( j\omega_0 -\frac{1}{\tau_l(t)} -\frac{1}{\tau_c(t)} \Big) a(t) - j \mu E_0 e^{j(\omega - \omega_r)t}\\
%    & E_{out}(t) = E_0 e^{j\omega_0 t} - j\mu a(t)
%\end{align}

\begin{align} 
    & \frac{d a(t)}{d t} = \Big( j\omega_0 -\frac{1}{\tau_l(t)} -\frac{1}{\tau_c(t)} \Big) a(t) - j \mu E_{in} \label{eq:ring_partial1}\\
    & E_{out}(t) = E_{in} - j\mu a(t) \label{eq:ring_partial2}
\end{align}

Here, $E_{in}=E_0 e^{j\omega_L t}$ in the input wave amplitude, typically a CW laser with wavelength, $\lambda_L$. $E_{out}$ is the output or through wave amplitude. The energy cross-coupling factor ($\mu$) relates to the power coupling coefficient ($\kappa^2$) and time-constant ($\tau_c$) as  \cite{little1997microring}

\begin{equation} \label{eq:mu_kappa}
    \mu^2 = \frac{\kappa^2 \nu_g}{2\pi R} = \frac{2}{\tau_c}
\end{equation}
where $\nu_g$ is the group velocity.  
%Also, here $E_0 e^{j(\omega - \omega_r)t}$ frequency-translated analytic model which significantly speeds up the differential equation simulation in Cadence Spectre. 
Solving Eqn. \ref{eq:ring_partial1} for steady-state zero-bias condition, $\tau_l(t)=\tau_{l}$ and $\tau_c(t)=\tau_{c}$, and input $E_{in}=E_0 e^{j\omega t}$, we obtain the energy amplitude \cite{rhim2015verilog}
\begin{equation} \label{eq:a_steady_state}
    a = \frac{-j \sqrt{\frac{2}{\tau_{c}}}}{j(\omega-\omega_0) + \frac{1}{\tau}} E_{in}
\end{equation}

Substituting this into Eqn. \ref{eq:ring_partial2}, we obtain the Lorentzian power transmission 
\begin{align} \label{eq:T_steady_state}
    T_{pwr} = \left| \frac{E_{out}}{E_{in}}\right|^2 = \frac{(\omega-\omega_0)^2 + \big(\frac{1}{\tau_l} - \frac{1}{\tau_c}\big)^2}{(\omega-\omega_0)^2 + \big(\frac{1}{\tau_l} + \frac{1}{\tau_c}\big)^2} 
\end{align}

\begin{figure}[!ht]
\centering
\scriptsize
\includegraphics[width=0.8\columnwidth]{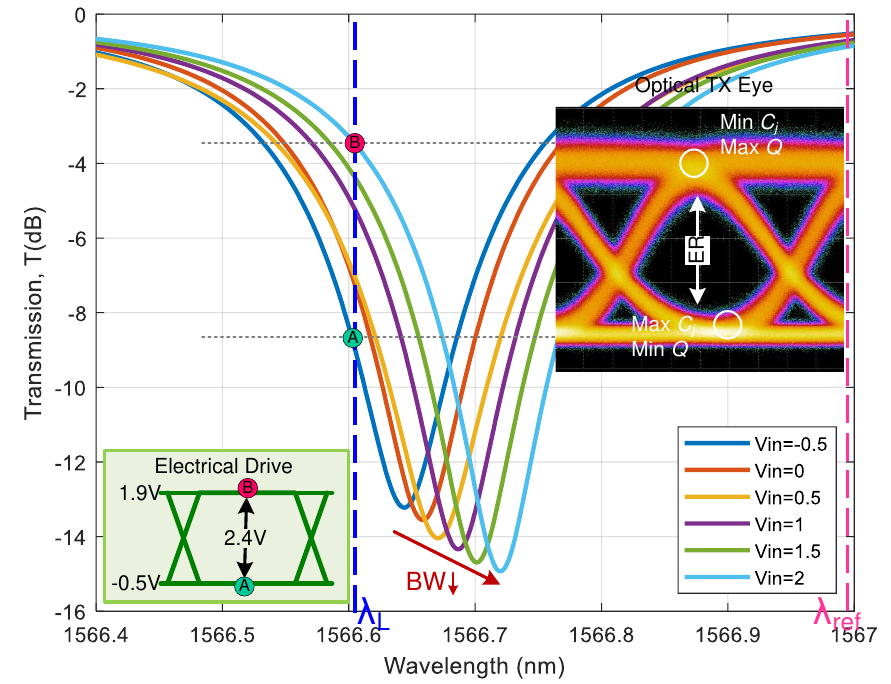}
\caption{Static transmission response of a microring modulator with wavelength on the x-axis. Here, the laser input is applied at $\lambda_{L}$ to obtain optical modulation. The model uses a reference wavelength, $\lambda_{R}$.}
\label{fig:static_response}
\end{figure}

\textbf{Eqn.} \ref{eq:T_steady_state} describes the static transmission characteristics of the resonant modulator and is plotted in \textbf{Fig.} \ref{fig:static_response}. As the MDM is modulated with the applied voltage, $v_{m}(t)$, it changes the ring's resonance frequency (or wavelength) as $\omega_0(v_m)=\frac{2\pi c}{\lambda_0(v_m)}$, and the time-constants $\tau_l(v_m)$ and $\tau_c(v_m)$. 

%%%%%%%%%%%%%%%%%%%%%%%%%%%%%%%%%%%%%%%%%%%%%%%%%%%%%%%%%%%%%%%%%%%%%%%%%%%%%%%%%%%%%%%%%%%%%%%%%%%%%%%%
\subsection{Prior Lumped MRM Models and Their Limitations}
The prior Verilog-A compact model by Rhim \cite{rhim2015verilog} used an explicit solution for \textbf{Eqn.} \ref{eq:ring_partial1} for an incident optical field $E_i(t)=E_0e^{j\omega t}$, as the sum of the steady-state and homogeneous (or transient) response. The model iteratively computes the response at discrete time steps. A clock source with the constraint determined the time steps in this model that the clock time-period, $\Delta t$, is very small for the model approximation to hold. Thus, the simulation time steps in the Cadence Spectre simulator are not adaptively set by the solver but by the clock source. The model in \cite{rhim2015verilog} uses $\delta t \sim$100 fs time-steps for simulation of 28 Gbps data rate corresponding to the 1550nm wavelength or 193.54 THz, leading to slow transient simulations. Also, the clock input is not characteristic of the physical ring and is not an accurate implementation. 

Another major limitation of this model was that the junction capacitance, $C_j$, was kept constant. As the reverse bias increases the junction capacitance, $C_j(v_m)$ decreases, leading to a sharp increase in the rising edge rate or higher instantaneous electrical bandwidth, $f_{elec}$. This is observed as the non-linear rising edge in the eye response in \textbf{Fig.} \ref{fig:static_response}, which is not at all captured by Rhim's model in \cite{rhim2015verilog}. Accurate modeling of the dispersion in optical eye response is critical at high-speed co-simulation, especially for the PAM4 transmitters \cite{li2020112,sharma2021silicon}. 

Small-signal models for MRM have been explored \cite{karimelahi2016ring} but do not capture the nonlinear dynamics. A compact model was proposed for carrier-injection (or forward-biased) MRM for operation up to 9 Gbps \cite{wang2016compact}. This model explicitly computed a variation of \textbf{Eqn}. \ref{eq:T_steady_state} while considering the effect of phase shift due to the modulation of the effective index, $n$, and loss. The model has a similar implementation as the composite model with associated challenges discussed earlier.     

%%%%%%%%%%%%%%%%%%%%%%%%%%%%%%%%%%%%%%%%%%%%%%%%%%%%%%%%%%%%%%%%%%%%%%%%%%%%%%%%%%%%%%%%%%%%%%%%%%%%%%%%
\section{Resonant Modulator Compact Model for Rapid Simulation}

In this section, we introduce a compact model that allows rapid simulation of resonant modulators. This is obtained using a combination of baseband equivalent modeling and continuous-time implementation of \textbf{Eqns.} \ref{eq:ring_partial1}\ and \ref{eq:ring_partial2} that doesn't require an undesirable clock input to the MRM model. Furthermore, we have incorporated MRM/MDM thermo-optic tuning into the compact model.

%%%%%%%%%%%%%%
\subsection{Analytic or Baseband Equivalent Modeling}
 Analytic or baseband equivalent modeling is employed to translate the modulated laser response centered around 1550nm or $f_{L} \sim 193$ THz to the baseband, \textit{i.e.}, near DC) \cite{shawon2020rapid}. This relaxes the transient simulation time steps using Spectre (or other SPICE-like) circuit simulator. 
 % The reference frequency, $f_{R}$, can be different from the laser frequency, as we will see later in the case of resonant circuits.
Here, $\tilde{E}(t)$ is the analytic field, which is in turn related to the complex baseband field  $E_{bb}(t)$ as $\tilde{E}(t)= e^{j \omega_{\text{ref}} t} E_{bb}(t)$.
For the resonant modulator, the baseband equivalent response can be described by the relations:
\begin{equation} \label{eqn:analytic_eqn2}
%h(t) = \Re (\tilde{h}(t))= \Re (e^{j \omega_R t} h_{bb}(t)); \quad
\tilde{h}(t)= e^{j \omega{R} t} h_{bb}(t); \quad
\tilde{H}(j\omega)= H_{bb}(j(\omega-\omega_{R}))
\end{equation}
where $h_{bb}(t)$ is the complex baseband equivalent impulse response of the resonator, $h(t)$, with respect to the reference frequency, $\omega_{R}$. $H_{bb}(j\omega)$ and $H(j\omega)$ are their respective frequency response \cite{shawon2020rapid}. 
Since Verilog-A modeling language doesn't support complex variables, our modeling framework represents the real and imaginary signals and filter coefficients as an optical bus $E$[0:1]. The Cartesian field is given by $\underbrace{E[0]}_{x}+j \underbrace{E[1]}_{y}$ and polar by $\underbrace{E[0]}_{r} \underbrace{e^{jE[1]}}_{e^{j\phi}}$ \cite{shawon2020rapid}. Using the analytic framework, the MRM/MDM resonance frequency is $\omega_0' \rightarrow \omega_0-\omega_R$, where $\omega_R$ (or $\lambda_R$) is the chosen reference frequency (or wavelength).

\subsection{Proposed Analytic Compact Model}

In our model, the differential equation in \textbf{Eqn.} \ref{eq:ring_partial1} is implemented in the Verilog-A model using the analytic modeling approach. The resulting offset frequency, $\Delta f = c ( \frac{1}{\lambda_{L}} - \frac{1}{\lambda_{R}}$), is used to excite the laser and is kept in the range under 10-50 GHz to speed up the simulations. Also, due to the absence of complex variables in Verilog-A, the variables are realized using coupled differential equations with real variables as shown in \textbf{Eqns.} \ref{eq:ring_partial_verilogA1} - \ref{eq:ring_partial_verilogA4} below. The input is $E_{in}=E_{ix}+jE_{iy}$, transmitted wave is $E_{out}=E_{ox}+jE_{oy}$, and the ring energy variable is $a=a_{x}+ja_{y}$. 

\begin{align} 
    \frac{d a_x(t)}{d t} &= -\omega_0' a_y -\frac{a_x}{\tau(t)} + \mu E_{iy} \label{eq:ring_partial_verilogA1} \\
    \frac{d a_y(t)}{d t} &= +\omega_0' a_x -\frac{a_y}{\tau(t)} - \mu E_{ix} \label{eq:ring_partial_verilogA2}
\end{align}

Also, the output fields are given as:
\begin{align} 
    E_{ox} &= E_{ix} + \mu a_y \label{eq:ring_partial_verilogA3} \\
    E_{oy} &= E_{iy} - \mu a_a \label{eq:ring_partial_verilogA4}
\end{align}

These equations are implemented using the Verilog-A \textbf{\textit{ddt}} operator and the \textbf{\textit{indirect branch statements}} as shown in the code \textbf{listing} \ref{MRM_code}. Employing the relatively obscure indirect branch statements ensures that the \textbf{Eqns.} \ref{eq:ring_partial_verilogA1} \& \ref{eq:ring_partial_verilogA2} are concurrently updated and key to the continuous-time implementation of the model differential equations. The equations are evaluated at the time-steps adaptively set by the Spectre solver being used and don't require any clock reference to compute the steady-state and transient solutions 
 to the differential equation. Thus, in the segments with less waveform change, the time steps will be determined by the solver, leading to overall faster simulation time while maintaining model accuracy.

\begin{lstlisting}[language=Verilog, basicstyle=\ttfamily\footnotesize, columns=fullflexible, breaklines=true, caption=Key sections of analytic MRM optical behavioral model., label=MRM_code] 

module Disk_modulator_tuned_va(in, out, N, P, HN, HP);
input [0:1] 		in;  // in[0] + j*in[1]
output [0:1] 		out;
input               N, P, HN, HP;

opticalField [0:1] 	in, out;
opticalField [0:1] 	a;
electrical          N, P, HN, HP;

.............................................
analog begin
	@(initial_step)	begin
		wref = 2*`PI*`P_C/lambda_ref; // Reference frequency for analytic representation		
	end 		// initial_step
	
	// Parameters curve-fitted to input voltage
	DLambda  = gamma*V(HP,HN)*I(HP,HN); // Microheater tuning term
    lambda0  = DLambda + lambda0_0 + lambda0_1*V(N,P);
	tau_c 	 = tau_c0 + tau_c1*V(N,P) + tau_c2*pow(V(N,P),2);
	tau_l 	 = tau_l0 + tau_l1*V(N,P) + tau_l2*pow(V(N,P),2);
	tau 	 = 1/( (1/tau_c) + (1/tau_l));
	mu 		 = sqrt(2/tau_c);
	w0 		 = 2*`PI*`P_C/lambda0 - wref;

    // Junction Capacitance
    Cj  = Cj0/pow((1+ V(N,P)/Vbi),mj);  // Depletion Capacitance
    I(N, P) <+ Cj*ddt(V(N,P));

    // Heater resistance model
    V(HP,HN) <+ Rheater*I(HP,HN);

	// Differential equation
	// ddt(a) = (j*w0 -1/tau)*a - j*mu*Ein
    OptE(a[0]) : ddt(OptE(a[0])) == -w0*OptE(a[1]) - OptE(a[0])/tau + mu*OptE(in[1]);
    OptE(a[1]) : ddt(OptE(a[1])) ==  w0*OptE(a[0]) - OptE(a[1])/tau - mu*OptE(in[0]);
	
	// Eout = Ein - j*mu*a
	OptE(out[0]) <+ OptE(in[0]) + mu*OptE(a[1]);
	OptE(out[1]) <+ OptE(in[1]) - mu*OptE(a[0]); 
end
endmodule




\end{lstlisting}

The thermo-optic tuning modeling fits the shift in resonance wavelength ($\lambda_0$) as a function of applied power to the microheater, $P_{h}=V_{h}I_{h}$, as in \textbf{Fig.} \ref{fig:MDM_model}. The model uses a constant heater resistance but can be easily modified to capture the nonlinear resistance, $R_h(P_h)$, due to self-heating at the cost of slower simulation.

\begin{figure}[!ht]
\centering
\scriptsize
\includegraphics[width=\linewidth]{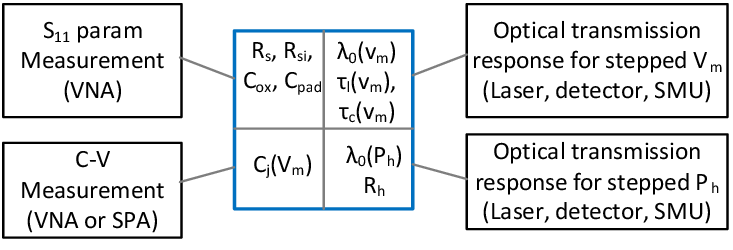}
\caption{Model parameter extraction from the experimental measurement results.}.
\label{fig:Model_flowchart}
\end{figure}

\subsection{Electrical Model Parameter Extraction} \label{sec:model_fitting}
\textbf{Fig.} \ref{fig:Model_flowchart} illustrates the set of measurements used to extract the parameters seen in the compact model in \textbf{Fig.} \ref{fig:MDM_model}. The model includes (i) the pn-junction capacitance, $C_j(v_m)$, with series resistance, $R_s$ (ii) substrate resistance, $R_{Si}$, and oxide capacitance, $C_{ox}$, and (iii) the parasitic pad capacitance, $C_{pad}$. A 50GHz Keysight PNA5225B VNA was used to obtain the MDM's $S_{11}$ response by probing its electrical pads. The electrical circuit parameters were fitted to the measured $S_{11}$ response at zero modulator bias voltage using Keysight ADS software \cite{pantouvaki201556gb,rhim2015verilog}. C-V characteristics were obtained by measuring $S_{11}$ response at -1V and 0.4V bias voltages and extracting $C_j$. Subsequently, the following \textit{pn}-junction capacitance model was fitted to these three data points 
\begin{equation}
    C_j(v_m) = \frac{C_{j0}}{\big(1+\frac{v_m}{V_{bi}} \big)^{m_j}}
\end{equation}

where $V_{bi}$ and $m_j$ are the built-in potential and grading exponent parameter respectively. The characteristics also can be obtained by measuring the C-V response using a semiconductor parameter analyzer (SPA). The extracted parameters of the electrical model for the fabricated MDM are shown in Table \ref{tab:electrical_param}. The calculations show an electrical self bandwidth, $f_{elec}$, around 18GHz, understandably due to the large zero-bias junction capacitance in the microdisk.

\begin{figure}[!ht]
\centering
\scriptsize
(a)
\includegraphics[width=0.5\linewidth]{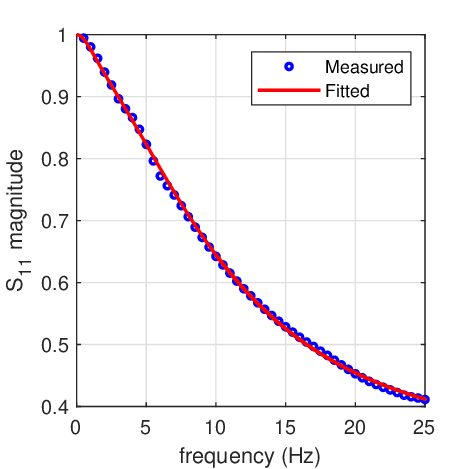} \\
(b)
\includegraphics[width=0.5\linewidth]{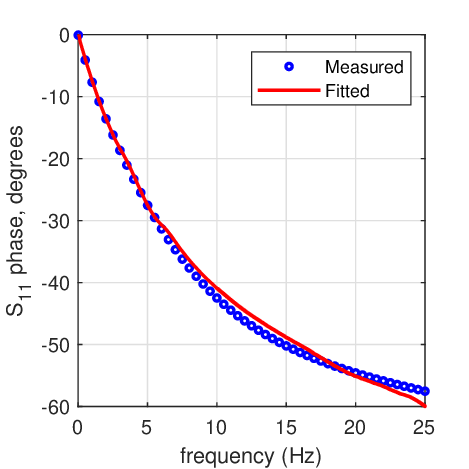} \\
(c)
\includegraphics[width=0.5\linewidth]{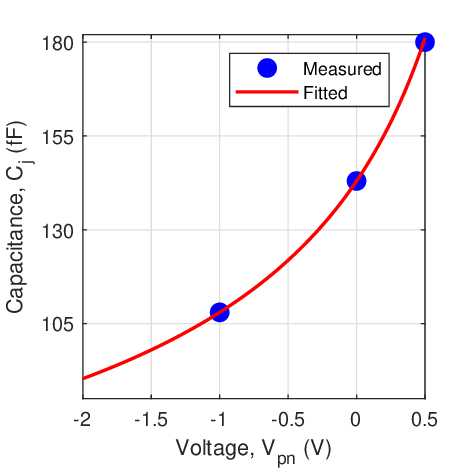}
\caption{Experimentally measured and fitted $S_{11}$ response of the MRM at zero bias: (a) magnitude, (b) phase. (c) Junction capacitance, $C_j$, fitted to the voltage, $V_{pn}$.}
\label{fig:S11fit}
\end{figure}

\begin{table}[ht]
\centering
\caption{Extracted electrical parameters for the MDM model seen in \textbf{Fig.} \ref{fig:MDM_model}.}
\label{tab:electrical_param}
\begin{tabular}{|c|c|c|c|c|c|c|}
\hline
$C_{j0}$   & $V_{bi}$  & $m_j$ & $R_s$  & $C_{ox}$  & $R_{Si}$ & $C_{pad}$ \\ \hline
143fF & 1.328V & 0.5  & 79.28$\Omega$ & 65.3fF & 1.4k$\Omega$ & 20.3fF \\ \hline\hline
$R_{h}$   & $\gamma$  & $\tau_h$ & & & & \\ \hline
8k$\Omega$    & 251 pm/mW  & 15$\mu$s & & & & \\ \hline

\end{tabular}
\end{table}

%These characteristics are experimentally extracted as the detuning of the Lorentzian curves. Also, the nonlinear resistance of the microheater is extracted using a source measurement unit (SMU) and built into the model.  

\subsection{Optical Model Parameter Extraction}
As discussed earlier in \textbf{Section} \ref{sec:optical_lumped_modeling}, the static and dynamic characteristics of the optical sub-model of the MRM/MDM compact model are captured by the three voltage-dependent parameters, $\omega_0(v)=\frac{1}{\lambda_0(v)}$, $\tau_c(v)$, and $\tau_l(v)$. The static response of the MDM is obtained using wavelength sweeps using the Keysight N7778C tunable laser and N7744A detector for several DC bias values ranging from -0.5V to 2.5V. The DC bias is applied using a Keysight B2901B source measurement unit (\textbf{SMU}). These voltage-dependent family of static transmission response are given by \textbf{Eqn.} \ref{eq:T_steady_state} and seen in \textbf{Fig.} \ref{fig:static_response}.   

\begin{figure}[!ht]
\centering
\scriptsize
(a)
\includegraphics[width=0.75\linewidth]{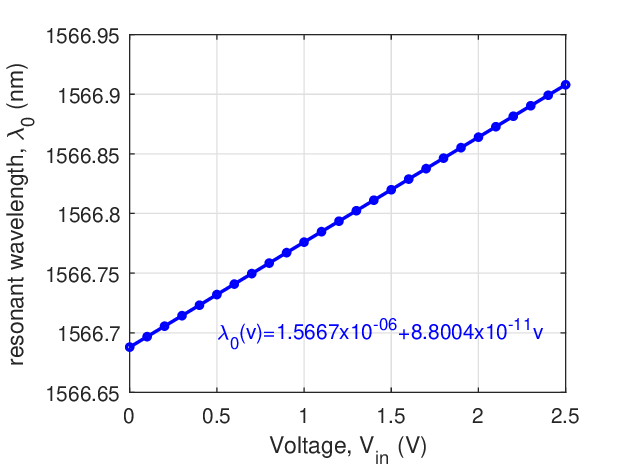}\\
(b)
\includegraphics[width=0.7\linewidth]{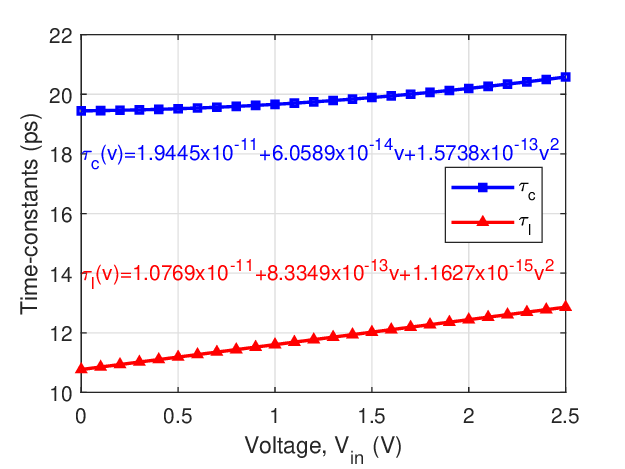}
\caption{Fitted model parameters with respect to the applied voltage for the MDM: (a) $\lambda_0$, and (b) $\tau_c$, and $\tau_l$.}
\label{fig:MRM_fitted_params} 
\end{figure}

The measured static responses are fitted to Lorentzian transmission to de-embed the grating coupler (\textbf{GC}) loss and the ripples in the optical response. First, the parameter $\lambda_0(v)$ is obtained by finding transmission magnitude mimima, $|T_0|$, at $\lambda_0$. The extracted $\lambda_0(v)$ is curve fitted to the voltage using Matlab function \textit{polyfit} as
\begin{equation}
    \lambda_0(v) \approx \Delta \Lambda+ \lambda_{0,0} + \lambda_{0,1} v + \lambda_{0,2} v^2
\end{equation}
where $\Delta \Lambda$ is the thermo-optic wavelength tuning term discussed earlier in this Section.

Next, we determine the curve-fits for the losses $\tau_c$, and $\tau_l$. Simultaneous nonlinear fitting of the experimentally measured transmission to a Lorentzian response with two variables, $\tau_c$, and $\tau_l$, leads to error-prone results. Instead, we reduce the optimization problem to a non-linear fit of a single variable. This is accomplished by using the minimum value of $T_0$, which provides the relation:

\begin{equation} \label{eq: T0_fitting1}
    T_0 = \left| \frac{\frac{1}{\tau_l}-\frac{1}{\tau_c}}{\frac{1}{\tau_l}+\frac{1}{\tau_c}} \right| >0 
\end{equation}

which using componendo-dividendo provides the constraint
\begin{equation} \label{eq: tau_fitting1}
    \frac{\tau_c}{\tau_l} = \frac{1+T_0}{1-T_0}
\end{equation}

This reduces the response to be fitted to a single variable, $\tau_l$

\begin{align} \label{eq: T_fitting2}
    T = \left| \frac{j(\omega-\omega_0) + \frac{a}{\tau_l}}{j(\omega-\omega_0) + \frac{b}{\tau_l} } \right| 
\end{align}

where $a=\frac{2T_0}{1+T_0}$ and $b=\frac{2}{1+T_0}$. \\

The Matlab function \textbf{\textit{lsqnonlin}} is used to curve-fit data seen in \textbf{Fig.} \ref{fig:static_response} to \textbf{Eqn.} \ref{eq: T_fitting2}. The resulting responses are fitted to the polynomial of the input reverse-bias modulation voltage \cite{rhim2015verilog}: 
\begin{align}
    \Delta \tau_c(v) &\approx \tau_{c0} + \tau_{c1} v + \tau_{c2} v^2\\
    \Delta \tau_l(v) &\approx \tau_{l0} + \tau_{l1} v + \tau_{cl} v^2
\end{align}

\textbf{Fig.} \ref{fig:MRM_fitted_params}  shows the curve-fitted parameters of the MDM used in this work. Here, we observe that as the MDM is depleted by applying a higher reverse bias, $V_{in}$, the resonance is red-shifted due to the increased effective index. We can observe the increase in the ring loss time-constant ($\tau_l$), suggesting that the optical losses are reduced due to the depletion of free carriers, and thus, the quality factor, $Q$, of the detuned disk progressively increases. This results in a corresponding decrease in the optical 3dB bandwidth, $f_{opt}=\frac{\omega_0}{2\pi Q}$, of the modulator. 

\begin{figure}[!ht]
\centering
\scriptsize
(a)
\includegraphics[width=0.7\linewidth]{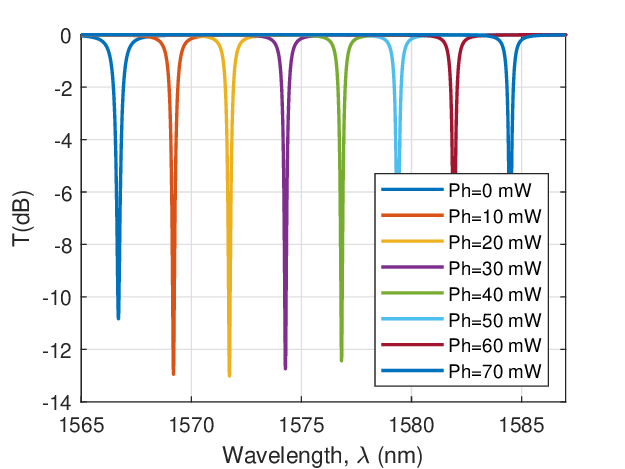}\\
(b)
\includegraphics[width=0.7\linewidth]{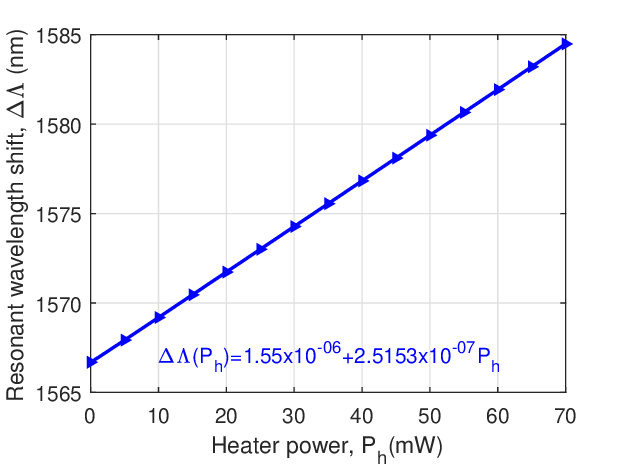}
\caption{(a) Thermo-optic tuning of the static transmission response for applied heater power. (b) Fitted $\Delta\Lambda$ parameter as a function of the applied heater power in the MDM.}
\label{fig:MDM_thermal_tuning} 
\end{figure}

Applying electrical power to the embedded heater produces an optical phase shift, $\Delta \phi \propto P_{h}$, which in turn manifests as a red shift in the resonant wavelength and is given by
\begin{equation}
     %\Delta \lambda_{0,0} (T) = \lambda_{1T} T + \lambda_{2T} T^2 + \cdots
      \Delta \Lambda = \gamma P_{h} = \gamma V_{h}I_{h}
\end{equation}

The swept wavelength transmission response of the MDM is obtained by applying a range of electrical power, $P_h$, in the microheater using an SMU (see \textbf{Fig.} \ref{fig:MDM_thermal_tuning} (a)). Using this characterization, the wavelength shift due to thermo-optic tuning, $\Delta\Lambda(P_h)$ is fitted to the heater power as shown in \textbf{Fig.} \ref{fig:MDM_thermal_tuning} (b).

\subsection{Compact Model Schematic}

\begin{figure}[!ht]
\centering
\scriptsize
\includegraphics[width=\linewidth]{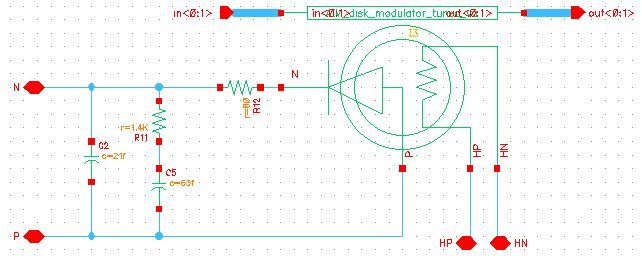}
%(b)
%\includegraphics[width=0.47\linewidth]{figures/schematic_top.png}
\caption{Compact model schematic of the MDM showing the electrical parasitics and the Verilog-A symbol.}
\label{fig:MRM_compact_model} 
\end{figure}

The MDM compact model schematic is shown in \textbf{Fig.} \ref{fig:MRM_compact_model}. Here, the electrical equivalent circuit is realized using passive components to improve simulation convergence, and the voltage-dependent junction capacitance is implemented inside the Verilog-A sub-model with its code seen in \textbf{Listing} 1. The details of the modeling framework are available in our previous work in \cite{shawon2019rapid, shawon2020rapid}.

\section{Experimental Results}

\subsection{Dynamic Transient Response}
\begin{figure}[!h]
\centering
\includegraphics[width=\columnwidth]{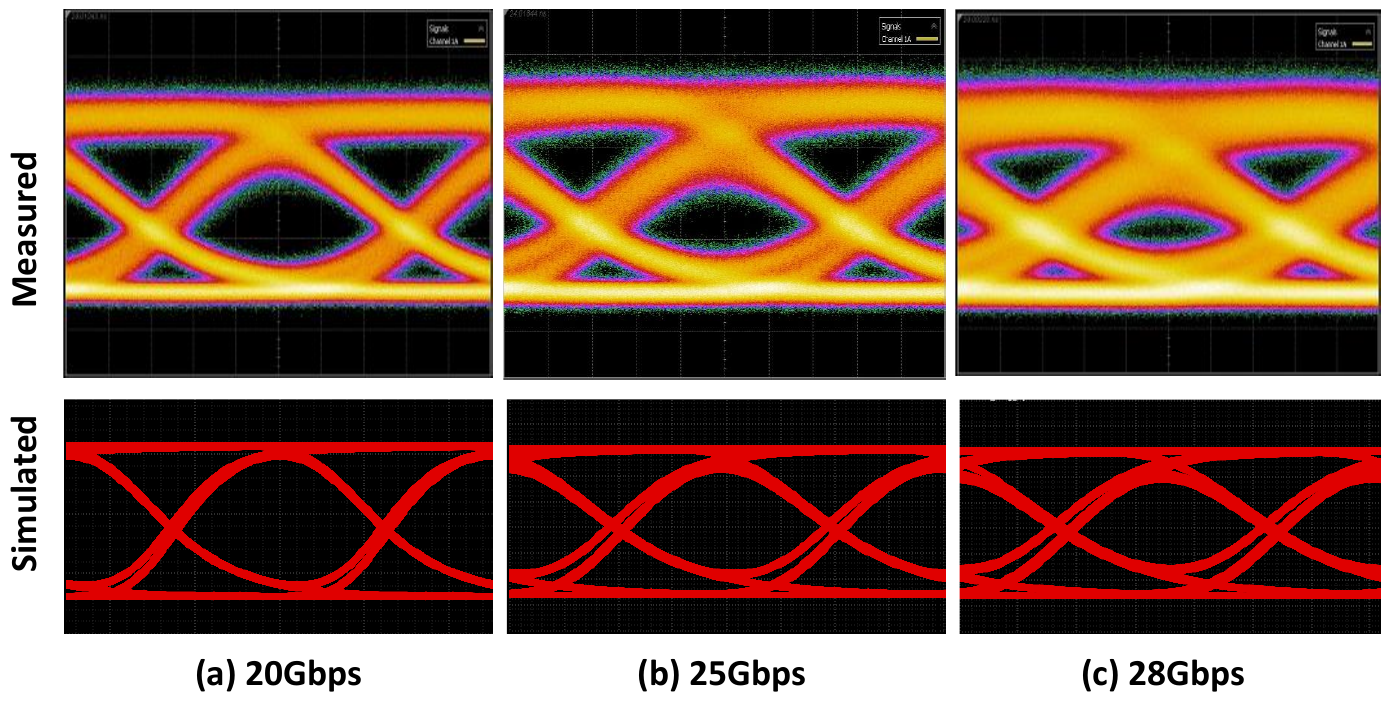}
\caption{Comparison of measured and simulated eye diagrams for the MDM driven by a 2$V_{pp}$ PRBS-13 NRZ data pattern at data rates: (a) 20 Gbps, (b) 25 Gbps, and (c) 28 Gbps. The laser wavelength is offset by -50pm with respect to the unbiased resonant wavelength.}
\label{fig:MDM_eye_comparison} 
\end{figure}

\textbf{Fig.} \ref{fig:MDM_eye_comparison} compares the measured and simulated eye diagrams for the modulator. The modulator's unbiased resonant wavelength, $\lambda_0(0)$, was around 1566.7nm. The input laser is applied at $\lambda_L=$1566.65nm, i.e., at an offset of -50pm. The modulator was driven from Keysight M8196A arbitrary waveform generator (AWG) with a 2$V_{pp}$ PRBS-13 non-return to zero (\textbf{NRZ}) pattern using an RF cable and an unterminated GS probe with $Z_0=$50$\Omega$ impedance. The eye diagrams were measured using the Keysight 86100D sampling scope with an optical module with 20GHz bandwidth. The simulated response from the compact model closely matches the experimental eye diagrams and accurately captures the dynamic nonlinearities. Here, we can see that the voltage-dependent electrical and optical bandwidths cause faster rise times and slower fall times with a long tail, respectively. It should be noted that the experimentally measured eyes exhibit higher jitter due to the resonant wavelength drift due to self-heating in the modulator. 

\begin{figure}[!h]
\centering
\includegraphics[width=\columnwidth]{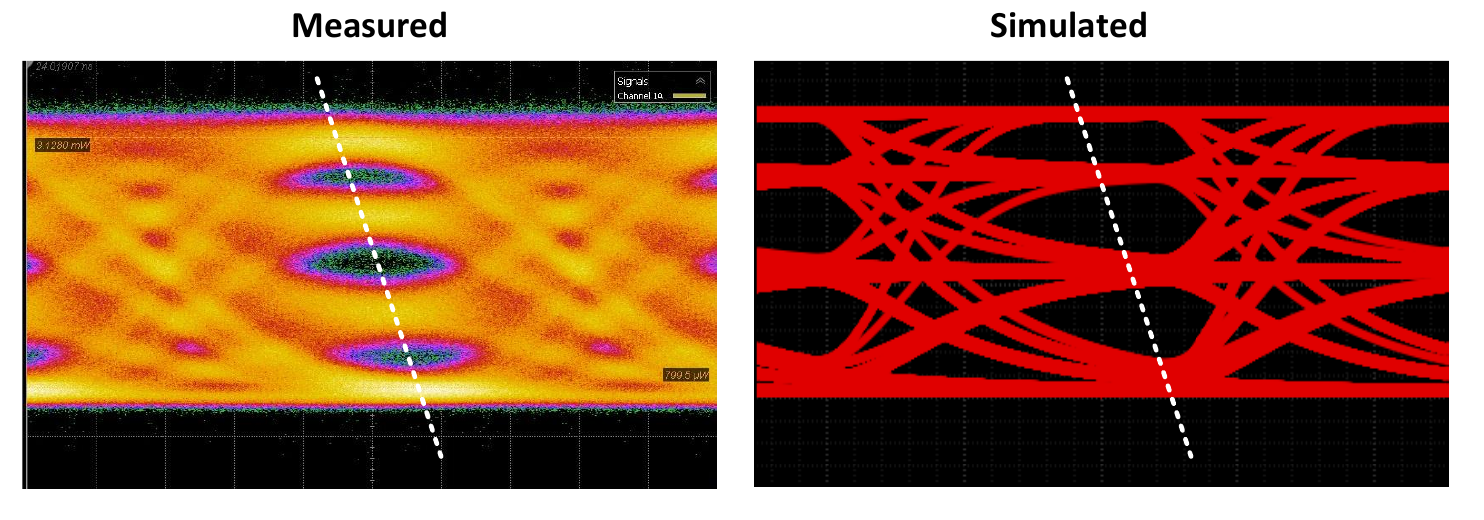}
\caption{Comparison of measured and simulated eye diagrams for the MDM driven by a 2$V_{pp}$ PAM4 data pattern at 20 GSps symbol rate. The laser wavelength is offset by -50pm with respect to the unbiased resonant wavelength. The dashed line shows the alignment of the centers of dispersed eye-opening.}
\label{fig:MDM_PAM4_eye_comparison} 
\end{figure}

\textbf{Fig.} \ref{fig:MDM_PAM4_eye_comparison} shows the same comparison for a 20GSps PAM4 signal. The model accurately captures the characteristic dispersion in the PAM4 eye response due to the dynamic nonlinearities. The model can closely predict the transmitter dispersion eye closure quarternary (TDECQ) metric, an important performance metric in optical PAM4 transmitters \cite{saxena2024ringTcas2}.

\subsection{Static Response and Thermo-optic Tuning}
The static transmission response for the MDM was also simulated to demonstrate the comprehensiveness of the developed compact model. \textbf{Fig.} \label{fig:MRR_static_resp}(a) shows the frequency response of the MDM for DC bias stepped from -0.5V to 2.5V, which was obtained using our analytic frequency chip method (\textbf{FCM}) detailed in \cite{shawon2019rapid}. Next, \textbf{Fig.} \label{fig:MRR_static_resp}(b) shows the thermal tuning of the MDM response using the FCM simulation as the heater power is increased in discrete steps. The simulated static response also matches the measurements.

\begin{figure}[!ht]
\centering
\scriptsize
(a)
\includegraphics[width=0.7\linewidth]{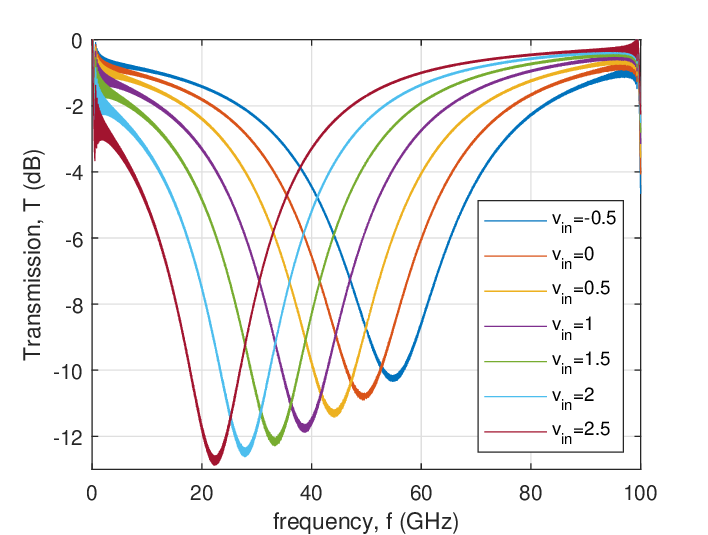}\\
(b)
\includegraphics[width=0.7\linewidth]{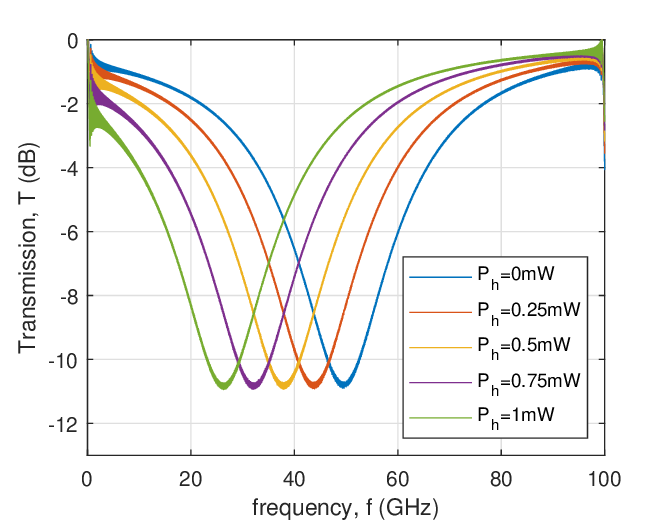}
\caption{Simulated static transmission response of the MDM using the analytic frequency chirp method (FCM) with Cadence Spectre: (a) modulator detuning as for a drive voltage from -0.5V to 2.5V. (b) Thermo-optic tuning as the heater power is stepped from 0 to 1mW. Note that the horizontal axis is the frequency and thus, the resonance detuning is mirrored to the left when compared to Fig. \ref{fig:static_response} as $f=\frac{c}{\lambda}$.}
\label{fig:MRR_static_resp} 
\end{figure}

\subsection{Comparison with Prior Art}

The simulations were performed using Cadence Virtuoso (IC618) and Spectre simulator on an Intel Xeon 4214R dual CPU server with 64GB memory. Both the models were simulated using transient analysis for the same number of unit intervals (\textbf{UI}s) at 28Gbps NRZ data rate. The proposed model was simulated for the same number of samples with a Spectre \textit{maxstep} of 100fs, which sets the accuracy for the Spectre transient simulation. Several transient runs were performed for both models, and the CPU time was recorded from the simulation log. Rhim's model, on average, took 84 seconds for 1000 UIs, while ours took 12 seconds.  For 3000 UIs, the average simulation time was 200 seconds and 35 seconds, respectively. Thus, the proposed compact model offers around 7$\times$ speed up compared to the prior art. Furthermore, Rhim's model only captures the nonlinear optical response (and not electrical) and produces jagged and discontinuous optical waveforms due to clock-based sampling. The proposed model accurately captures electrical and optical dynamic nonlinearity and produces continuous waveforms. 

\section{Conclusion}
We presented a novel and comprehensive resonant modulator compact model that can be used to model silicon photonic microdisk and microring modulators. The model captures the dynamic effects of the MDM more accurately than the prior art by considering both electrical and optical voltage-dependent nonlinearities. Also, thanks to a novel Verilog-A realization using real-valued coupled differential equations and analytic modeling, the model offers around 7$\times$ speed up in transient simulation time. Lastly, the mode integrates the thermo-optic tuning of the resonant wavelength. Simulated results show good agreement with measurement data and accurately simulate the NRZ and PAM4 eye response. The model will form a part of the essential toolkit for hybrid electronic-photonic circuit co-simulations of PICs using resonant modulators.

\section*{Funding}
This work was supported by the Air Force Office of Sponsored Research (AFOSR) YIP Award FA9550-17-1-0076,  NSF CAREER Award EECS-2014109, and DARPA YFA Award HR00112110001.

\section*{Contributions}
VS conceived and developed the compact model, fitted the parameters, performed simulations, and composed the entire manuscript. MJS taped out the chip test structures and performed experimental measurements and S11 response fitting. VS supervised the overall project.

\section*{Acknowledgement}
The authors thank Shubham Mishra for assistance with simulating some of the test benches. The authors also thank AIM Photonics and gratefully acknowledge APSUNY PDK Component Library.

%%%%%%%%%%%%%%%%%%%%%%% References %%%%%%%%%%%%%%%%%%%%%%%%%

%%%%%%%%%% If using BibTeX:
% \bibliography{bibtex/SiP_References_2021}
\bibliographystyle{IEEEtran}
\bibliography{bibtex/SiP_References_2023B}

\begin{thebibliography}{10}
\providecommand{\url}[1]{#1}
\csname url@samestyle\endcsname
\providecommand{\newblock}{\relax}
\providecommand{\bibinfo}[2]{#2}
\providecommand{\BIBentrySTDinterwordspacing}{\spaceskip=0pt\relax}
\providecommand{\BIBentryALTinterwordstretchfactor}{4}
\providecommand{\BIBentryALTinterwordspacing}{\spaceskip=\fontdimen2\font plus
\BIBentryALTinterwordstretchfactor\fontdimen3\font minus
  \fontdimen4\font\relax}
\providecommand{\BIBforeignlanguage}[2]{{%
\expandafter\ifx\csname l@#1\endcsname\relax
\typeout{** WARNING: IEEEtran.bst: No hyphenation pattern has been}%
\typeout{** loaded for the language `#1'. Using the pattern for}%
\typeout{** the default language instead.}%
\else
\language=\csname l@#1\endcsname
\fi
#2}}
\providecommand{\BIBdecl}{\relax}
\BIBdecl

\bibitem{li201525}
H.~Li, Z.~Xuan, A.~Titriku, C.~Li, K.~Yu, B.~Wang, A.~Shafik, N.~Qi, Y.~Liu,
  R.~Ding \emph{et~al.}, ``{A 25 Gb/s, 4.4 V-swing, AC-coupled ring
  modulator-based WDM transmitter with wavelength stabilization in 65 nm
  CMOS},'' \emph{IEEE Journal of Solid-State Circuits}, vol.~50, no.~12, pp.
  3145--3159, 2015.

\bibitem{wade2021monolithic}
M.~Wade, D.~Jeong, B.~Kim, M.~Zhang, W.~Bae, C.~Zhang, P.~Bhargava,
  D.~Van~Orden, S.~Ardalan, C.~Ramamurthy \emph{et~al.}, ``{Monolithic
  microring-based WDM optical I/O for heterogeneous computing},'' in \emph{2021
  Symposium on VLSI Circuits}.\hskip 1em plus 0.5em minus 0.4em\relax IEEE,
  2021, pp. 1--2.

\bibitem{saxena2024ringTcas2}
V.~Saxena, A.~Kumar, S.~Mishra, S.~Palermo, and K.~R. Lakshmikumar, ``{Optical
  Interconnects Using Hybrid Integration of CMOS and Silicon-Photonic ICs},''
  \emph{IEEE Transactions on Circuits and Systems II: Express Briefs (Early
  Access)}, 2024.

\bibitem{rizzo2022petabit}
A.~Rizzo, S.~Daudlin, A.~Novick, A.~James, V.~Gopal, V.~Murthy, Q.~Cheng, B.~Y.
  Kim, X.~Ji, Y.~Okawachi \emph{et~al.}, ``Petabit-scale silicon photonic
  interconnects with integrated kerr frequency combs,'' \emph{IEEE Journal of
  Selected Topics in Quantum Electronics}, vol.~29, no. 1: Nonlinear Integrated
  Photonics, pp. 1--20, 2022.

\bibitem{huang2022prospects}
C.~Huang, V.~J. Sorger, M.~Miscuglio, M.~Al-Qadasi, A.~Mukherjee, L.~Lampe,
  M.~Nichols, A.~N. Tait, T.~Ferreira~de Lima, B.~A. Marquez \emph{et~al.},
  ``Prospects and applications of photonic neural networks,'' \emph{Advances in
  Physics: X}, vol.~7, no.~1, p. 1981155, 2022.

\bibitem{shawon2023silicon}
M.~J. Shawon and V.~Saxena, ``A silicon photonic reconfigurable optical analog
  processor (siroap) with a 4x4 optical mesh,'' in \emph{2023 IEEE
  International Solid-State Circuits Conference (ISSCC)}.\hskip 1em plus 0.5em
  minus 0.4em\relax IEEE, 2023, pp. 222--224.

\bibitem{ding2014compact}
R.~Ding, Y.~Liu, Q.~Li, Z.~Xuan, Y.~Ma, Y.~Yang, A.~E.-J. Lim, G.-Q. Lo,
  K.~Bergman, T.~Baehr-Jones \emph{et~al.}, ``{A compact low-power 320-Gb/s WDM
  transmitter based on silicon microrings},'' \emph{IEEE Photonics Journal},
  vol.~6, no.~3, pp. 1--8, 2014.

\bibitem{sharma2021silicon}
J.~Sharma, Z.~Xuan, H.~Li, T.~Kim, R.~Kumar, M.~N. Sakib, C.-M. Hsu, C.~Ma,
  H.~Rong, G.~Balamurugan \emph{et~al.}, ``{Silicon photonic microring-based
  4$\times$ 112 Gb/s WDM transmitter with photocurrent-based thermal control in
  28-nm CMOS},'' \emph{IEEE Journal of Solid-State Circuits}, vol.~57, no.~4,
  pp. 1187--1198, 2021.

\bibitem{mishra2022hybrid}
S.~Mishra, M.~J. Shawon, A.~Dorzhigulov, and V.~Saxena, ``{A Hybrid CMOS
  Photonic 25Gbps Microring Transmitter with a-0.5--1.2 V Direct-Coupled
  Drive},'' in \emph{2022 IEEE International Symposium on Circuits and Systems
  (ISCAS)}.\hskip 1em plus 0.5em minus 0.4em\relax IEEE, 2022, pp. 1000--1004.

\bibitem{kehan_MW2014}
K.~Zhu, V.~Saxena, and W.~Kuang, ``{Compact Verilog-A Modeling of Silicon
  Traveling-Wave Modulator for Hybrid CMOS Photonic Circuit Design},'' in
  \emph{57th IEEE International Midwest Symposium on Circuits and Systems
  (MWSCAS)}, Aug 2014.

\bibitem{zhuTCAS2015}
K.~Zhu, V.~Saxena, X.~Wu, and W.~Kuang, ``{Design Considerations for
  Traveling-Wave Modulator-Based CMOS Photonic Transmitters},'' \emph{IEEE
  Transactions on Circuits and Systems II: Express Briefs}, vol.~62, no.~4, pp.
  412--416, April 2015.

\bibitem{bcicts2018}
C.~Li, K.~Yu, J.~Rhim, K.~Zhu, N.~Qi, V.~Saxena, M.~Fiorentino, and S.~Palermo,
  ``{A 3D-Integrated 56 Gb/s NRZ/PAM4 Reconfigurable Segmented Mach-Zehnder
  Modulator based Si-photonics Transmitter},'' in \emph{IEEE BiCMOS and
  Compound Semiconductor Integrated Circuits and Technology Symposium
  (BCICTS)}, 2018.

\bibitem{shawon2020rapid}
M.~J. Shawon and V.~Saxena, ``Rapid simulation of photonic integrated circuits
  using verilog-a compact models,'' \emph{IEEE Transactions on Circuits and
  Systems I: Regular Papers}, 2020.

\bibitem{shawon2022JLT}
------, ``Fully automatic in-situ reconfiguration of rf photonic filters in a
  cmos-compatible silicon photonic process,'' \emph{IEEE Journal of Lightwave
  Technology}, 2022.

\bibitem{shawon2023automatic}
------, ``Automatic in-situ optical linearization of silicon photonic
  ring-assisted mz modulator for integrated rf photonic socs,'' in \emph{2023
  Optical Fiber Communications Conference and Exhibition (OFC)}.\hskip 1em plus
  0.5em minus 0.4em\relax IEEE, 2023, pp. 1--3.

\bibitem{zhu2016modeling}
K.~Zhu, C.~Li, N.~Qi, K.~Yu, M.~Fiorentino, R.~Beausoleil, and V.~Saxena,
  ``{Modeling of MZM-based Photonic Link Power Budget},'' in \emph{IEEE Optical
  Interconnects Conference (OIC), 2016}.\hskip 1em plus 0.5em minus 0.4em\relax
  IEEE, 2016, pp. 58--59.

\bibitem{sorace2015electro}
C.~Sorace-Agaskar, J.~Leu, M.~R. Watts, and V.~Stojanovic, ``{Electro-optical
  co-simulation for integrated CMOS photonic circuits with Verilog-A},''
  \emph{Optics express}, vol.~23, no.~21, pp. 27\,180--27\,203, 2015.

\bibitem{rhim2015verilog}
J.~Rhim, Y.~Ban, B.-M. Yu, J.-M. Lee, and W.-Y. Choi, ``{Verilog-A behavioral
  model for resonance-modulated silicon micro-ring modulator},'' \emph{Optics
  express}, vol.~23, no.~7, pp. 8762--8772, 2015.

\bibitem{wang2016compact}
B.~Wang, C.~Li, C.-H. Chen, K.~Yu, M.~Fiorentino, R.~G. Beausoleil, and
  S.~Palermo, ``A compact verilog-a model of silicon carrier-injection ring
  modulators for optical interconnect transceiver circuit design,''
  \emph{Journal of Lightwave Technology}, vol.~34, no.~12, pp. 2996--3005,
  2016.

\bibitem{fahrenkopf2019aim}
N.~M. Fahrenkopf, C.~McDonough, G.~L. Leake, Z.~Su, E.~Timurdogan, and D.~D.
  Coolbaugh, ``The aim photonics mpw: A highly accessible cutting edge
  technology for rapid prototyping of photonic integrated circuits,''
  \emph{IEEE Journal of Selected Topics in Quantum Electronics}, vol.~25,
  no.~5, pp. 1--6, 2019.

\bibitem{baehr201225}
T.~Baehr-Jones, R.~Ding, A.~Ayazi, T.~Pinguet, M.~Streshinsky, N.~Harris,
  J.~Li, L.~He, M.~Gould, Y.~Zhang \emph{et~al.}, ``A 25 gb/s silicon photonics
  platform,'' \emph{arXiv preprint arXiv:1203.0767}, 2012.

\bibitem{pantouvaki201556gb}
M.~Pantouvaki, P.~Verheyen, J.~De~Coster, G.~Lepage, P.~Absil, and
  J.~Van~Campenhout, ``56gb/s ring modulator on a 300mm silicon photonics
  platform,'' in \emph{2015 European Conference on Optical Communication
  (ECOC)}.\hskip 1em plus 0.5em minus 0.4em\relax IEEE, 2015, pp. 1--3.

\bibitem{timurdogan2013vertical}
E.~Timurdogan, C.~M. Sorace-Agaskar, E.~S. Hosseini, G.~Leake, D.~D. Coolbaugh,
  and M.~R. Watts, ``Vertical junction silicon microdisk modulator with
  integrated thermal tuner,'' in \emph{CLEO: 2013}.\hskip 1em plus 0.5em minus
  0.4em\relax IEEE, 2013, pp. 1--2.

\bibitem{timurdogan2014ultralow}
E.~Timurdogan, C.~M. Sorace-Agaskar, J.~Sun, E.~Shah~Hosseini, A.~Biberman, and
  M.~R. Watts, ``An ultralow power athermal silicon modulator,'' \emph{Nature
  communications}, vol.~5, no.~1, pp. 1--11, 2014.

\bibitem{li202012}
H.~Li, G.~Balamurugan, M.~Sakib, R.~Kumar, H.~Jayatilleka, H.~Rong, J.~Jaussi,
  and B.~Casper, ``A 3d-integrated microring-based 112gb/s pam-4
  silicon-photonic transmitter with integrated nonlinear equalization and
  thermal control,'' in \emph{2020 IEEE International Solid-State Circuits
  Conference-(ISSCC)}.\hskip 1em plus 0.5em minus 0.4em\relax IEEE, 2020, pp.
  208--210.

\bibitem{kumar2023power}
V.~Kumar, S.~Mishra, and V.~Saxena, ``{Power Linear DACs (PLDACs) for
  Configuration and Control of Silicon Photonic Integrated Circuits},'' in
  \emph{2023 IEEE International Symposium on Circuits and Systems
  (ISCAS)}.\hskip 1em plus 0.5em minus 0.4em\relax IEEE, 2023, pp. 1--5.

\bibitem{RingModLumerical}
\BIBentryALTinterwordspacing
{INTERCONNECT: Enabling time and frequency domain simulation of photonic
  integrated circuits with microring modulators}. [Online]. Available:
  \url{https://www.lumerical.com/learn/whitepapers/interconnect-enabling-time-and-frequency-domain-simulation-of-photonic-integrated-circuits-with-microring-modulators/}
\BIBentrySTDinterwordspacing

\bibitem{shawon2019rapid}
M.~J. Shawon and V.~Saxena, ``Rapid simulation of photonic integrated circuits
  using verilog-a compact models,'' in \emph{2019 IEEE 62nd International
  Midwest Symposium on Circuits and Systems (MWSCAS)}.\hskip 1em plus 0.5em
  minus 0.4em\relax IEEE, 2019, pp. 424--427.

\bibitem{zhang2010silicon}
L.~Zhang, Y.~Li, J.-Y. Yang, M.~Song, R.~G. Beausoleil, and A.~E. Willner,
  ``{Silicon-based Microring Resonator Modulators for Intensity Modulation},''
  \emph{{IEEE Journal of Selected Topics in Quantum Electronics}}, vol.~16,
  no.~1, pp. 149--158, 2010.

\bibitem{little1997microring}
B.~E. Little, S.~T. Chu, H.~A. Haus, J.~Foresi, and J.-P. Laine, ``Microring
  resonator channel dropping filters,'' \emph{Journal of lightwave technology},
  vol.~15, no.~6, pp. 998--1005, 1997.

\bibitem{li2020112}
H.~Li, G.~Balamurugan, M.~Sakib, J.~Sun, J.~Driscoll, R.~Kumar, H.~Jayatilleka,
  H.~Rong, J.~Jaussi, and B.~Casper, ``A 112 gb/s pam4 silicon photonics
  transmitter with microring modulator and cmos driver,'' \emph{Journal of
  Lightwave Technology}, vol.~38, no.~1, pp. 131--138, 2020.

\bibitem{karimelahi2016ring}
S.~Karimelahi and A.~Sheikholeslami, ``Ring modulator small-signal response
  analysis based on pole-zero representation,'' \emph{Optics Express}, vol.~24,
  no.~7, pp. 7585--7599, 2016.

\end{thebibliography}

%%%%%%%%%% If preparing manually:
% \newcommand{\enquote}[1]{``#1''}

% \bibitem{Zhang:14}
% Y.~Zhang, S.~Qiao, L.~Sun, Q.~W. Shi, W.~Huang, L.~Li, and Z.~Yang,
%   \enquote{Photoinduced active terahertz metamaterials with nanostructured
%   vanadium dioxide film deposited by sol-gel method,}
%   {\protect\JournalTitle{Optics Express}} \textbf{22}, 11070--11078 (2014).

% {Optical Society}, \enquote{{OSA Publishing},}
%   \url{http://www.osapublishing.org}.

% \bibitem{FORSTER2007}
% P.~Forster, V.~Ramaswamy, P.~Artaxo, T.~Bernsten, R.~Betts, D.~Fahey,
%   J.~Haywood, J.~Lean, D.~Lowe, G.~Myhre, J.~Nganga, R.~Prinn, G.~Raga,
%   M.~Schulz, and R.~V. Dorland, \enquote{Changes in atmospheric consituents and
%   in radiative forcing,} in \enquote{Climate Change 2007: The Physical Science
%   Basis. Contribution of Working Group 1 to the Fourth assesment report of
%   Intergovernmental Panel on Climate Change,}  S.~Solomon, D.~Qin, M.~Manning,
%   Z.~Chen, M.~Marquis, K.~B. Averyt, M.~Tignor, and H.~L. Miler, eds.
%   (Cambridge University Press, 2007).

% \end{thebibliography}

\end{document}